\documentclass{llncs}          
\pdfoutput=1
\usepackage{listings}
\usepackage{graphics,epsfig}
\usepackage{amssymb}
\usepackage{stmaryrd}
\usepackage{subfigure}
\usepackage{color}
\usepackage{tikz}
\usetikzlibrary{arrows,automata,shapes}
\usetikzlibrary{backgrounds,fit,decorations.pathreplacing}

\lstdefinelanguage{rcos} {alsoletter={\#, \&},
  mathescape=true,
  basicstyle=\small,
  escapechar=\@,
  boxpos=c,
  morekeywords= {theorem, lemma, apply, by, constdefs, definition, where,
    infixl, types, consts, primrec, have, from, show, let,  proof, qed, is,
    sorry, with, assume, fix, thus, hence, datatype, if, then, else, in, case,
    of, var, end, class, public
  },
  emphstyle={\bf},
  emph={[2]int, char, string, self, boolean},
  emphstyle={[2]\it},
literate=
  {->}{{$\rightarrow$}}2
  {=>}{{$\Rightarrow$}}2
  {-->}{{$\Rightarrow$}}2
  {\\forall}{{$\forall$}}2
  {\\exist}{{$\exists$}}2
  {\\exists}{{$\exists$}}2
  {AND}{{$\& \&$}}3
  {ALL}{{$\forall$}}2
  {EX}{{$\exists$}}2
  {\%}{{$\lambda$}}1
  {\\\/}{{$\sqcap$}}2
  {|-}{{$\vdash$}}2
  {==}{{$\equiv$}}2
  {==>}{{$\Rightarrow$}}2
  {~}{{$\neg$}}1
  {~=}{{$\neq$}}2
  {:=}{{$:=$}}2
  {;}{{\tt ;}}1
}

\lstdefinelanguage{isa} {alsoletter={\#, \&},
  mathescape=true,
  basicstyle=\small,
  escapechar=\@,
  boxpos=c,
  morekeywords= {theorem, lemma, apply, by, constdefs, definition, where,
    infixl, types, consts, primrec, have, from, show, let,  proof, qed, is,
    sorry, with, assume, fix, thus, hence, datatype, if, then, else, in, case,
    of
  },
  emph={[2]int, char, string, self, boolean},
  emphstyle={[2]\it},
literate=
  {->}{{$\rightarrow$}}2
  {=>}{{$\Rightarrow$}}2
  {-->}{{$\Rightarrow$}}2
  {\\forall}{{$\forall$}}2
  {\\exist}{{$\exists$}}2
  {\\exists}{{$\exists$}}2
  {AND}{{$\& \&$}}3
  {ALL}{{$\forall$}}2
  {EX}{{$\exists$}}2
  {\%}{{$\lambda$}}1
  {\\\/}{{$\sqcap$}}2
  {|-}{{$\vdash$}}2
  {==}{{$\equiv$}}2
  {==>}{{$\Rightarrow$}}2
  {~}{{$\neg$}}1
  {~=}{{$\neq$}}2
  {:=}{{\tt :=}}2
  {;}{{\tt ;}}1
  { graph }{{ $G$ }}2
}

\lstnewenvironment{isaenv} {
  \lstset{extendedchars=true,columns=flexible}
  \lstset{language=isa,deletestring=[b]', basicstyle=\footnotesize}%
} {}

\lstnewenvironment{rcosenv} {
  \lstset{mathescape=true}
  \lstset{extendedchars=true,columns=flexible}
  \lstset{language=rcos, deletestring=[b]', basicstyle=\footnotesize \sf}%
  \lstset{escapechar=\@}
} {}

\lstdefinelanguage{maude} {alsoletter={\#, \&},
  mathescape=true,
  basicstyle=\small,
  escapechar=\@,
  boxpos=c,
  morekeywords= {theorem, lemma, apply, by, constdefs, definition, where,
    infixl, types, consts, primrec, have, from, show, let,  proof, qed, is,
    sorry, with, assume, fix, thus, hence, datatype, if, then, else, in, case,
    of
  },
  emph={[2]int, char, string, self, boolean},
  emphstyle={[2]\it},
literate=
  {~>}{{$\sim>$}}2
}

\lstnewenvironment{maudeenv} {
  \lstset{extendedchars=true,columns=flexible}
  \lstset{language=maude,deletestring=[b]', basicstyle=\footnotesize}%
} {}

\newcommand{\isa}[1]{{\lstinline[language=isa, basicstyle={\footnotesize \sf} ]@#1@}}
\newcommand{\maude}[1]{{\lstinline[language=maude, basicstyle={\footnotesize \sf} ]+#1+}}

\newcommand{\rcos}[1]{{\lstinline[language=rcos, basicstyle={\footnotesize \sf} ]^#1^}}

\newcommand{\chaos}{\ensuremath{\textsc{chaos}}}
\newcommand{\myskip}{\ensuremath{\textsc{skip}}}

\newcommand{\pdo}{\ensuremath{\textsf{do }}}

\usepackage{mathptmx}      
\usepackage{fancyhdr}

\begin{document}

\mainmatter

\title{A Framework for Automated and Certified Refinement Steps\thanks{This work has been supported by the project GAVES of the Macao S\&TD Fund, the 973 program 2009CB320702, STCSM 08510700300, the projects NSFC-60970031,  NSFC-91018012 and NSFC-61100061, the EU FP7-ICT project NESSoS (Network of Excellence on Engineering Secure Future Internet Software Services and Systems) under the grant agreement n. 256980, and the EU FP7-PEOPLE project DiVerMAS (Distributed System Verification with MAS-based Model Checking) under the grant agreement n. 252184.}
}

\titlerunning{A Framework for Automated and Certified Refinement Steps}        

\author{Andreas Griesmayer\inst{1}         \and
        Zhiming Liu\inst{2}                \and
        Charles Morisset\inst{3}           \and
        Shuling Wang\inst{4} 
}

\authorrunning{A. Griesmayer, Z. Liu, C. Morisset and S. Wang} 

\institute{Imperial College, London, UK \\
  \email{andreas.griesmayer@imperial.ac.uk} \\
  \and
  IIST, United Nations University, Macao\\
  \email{lzm@iist.unu.edu}\\
  \and
  Security Group, IIT - CNR, Italy\\
  \email{charles.morisset@cnr.iit.it} \\
  \and
  State Key Lab. of Computer Science\\
  Institute of Software, Chinese Academy of Sciences\\
  \email{wangsl@ios.ac.cn}\\
}


\maketitle

\begin{abstract}
    The refinement calculus provides a methodology for
    transforming an abstract specification into a concrete implementation,
    by following a succession of refinement rules. These rules have been
    mechanized in theorem-provers, thus providing a formal and rigorous
    way to prove that a given program refines another one. In a previous
    work, we have extended this mechanization for object-oriented programs,
    where the memory is represented as a graph, and we have integrated our
    approach within the rCOS tool, a model-driven software development
    tool providing a refinement language. Hence, for any refinement step,
    the tool automatically generates the corresponding proof obligations
    and the user can manually discharge them, using a provided library of
    refinement lemmas. In this work, we propose an approach to automate
    the search of possible refinement rules from a program to another,
    using the rewriting tool Maude. Each refinement rule in Maude is
    associated with the corresponding lemma in Isabelle, thus allowing the
    tool to automatically generate the Isabelle proof when a refinement
    rule can be automatically found. The user can add a new refinement
    rule by providing the corresponding Maude rule and Isabelle lemma.
\end{abstract}

\section{Introduction}

Software verification is about demonstrating that an
{\em implementation} (executable code) of the software
meets its {\em specification}
(formal description of the behavior) and several techniques, roughly classified into two categories, are available
in order to achieve this goal. The first category includes verification techniques taking both the specification and the implementation as inputs, and investigating whether the latter is an instantiation of the former. This investigation can be done for instance by running several instances of the implementation on specific inputs and monitoring whether the specification is respected during the execution ({\em e.g.} testing); by building an abstract model of the implementation and showing that the specification holds for any possible run ({\em e.g.} model-checking); by formally encoding both the specification and the implementation in a common language, using their respective semantics, and proving that the implementation logically implies the specification ({\em e.g.} theorem-proving).

The second category of verification techniques includes techniques where the implementation is generated from the specification, following a methodology guaranteeing by construction that the implementation meets the specification.  A typical approach is the program extraction from proofs~\cite{DBLP:conf/lcc/BergerS94}, which, using the Curry-Howard correspondence, where, from the proof of existence of some input satisfying a specification, a program implementing this specification can be automatically extracted. For instance, Coq's extraction mechanism of a program from a formal specification~\cite{DBLP:conf/types/Letouzey02} certifies that the program is correct with respect to this specification. Model-Driven Engineering~\cite{Kent:2002:MDE:647983.743552} considers a program as a model, which can be derived from another one using model transformations~\cite{Mens2006125}. Similarly, the refinement calculus~\cite{Back78,Morgan94} provides a formal language in which both abstract specifications and concrete implementations can be expressed and mixed, and some formal refinement rules describing how to transform a program into another, more concrete one.

In general, these techniques offer equivalent results: if an implementation is proved to meet a specification, then there exists a refinement chain from the specification to the implementation, and conversely, if there exists a chain of refinement from a specification to an implementation, then it can be proven that this implementation satisfies this specification. Moreover, different techniques can be combined in order to exploit the strengths of each technique for a given context. For instance, a model-checker can be integrated into the Isabelle theorem-prover~\cite{springerlink:10.1007/1159119127}, and test-cases can be automatically generated from an Isabelle specification~\cite{brucker.ea:hol-testgen:2009}. Similarly, model-checking can be used to verify refinement steps~\cite{springerlink:10.1007/11901433_38}, and the refinement calculus has been encoded into a theorem-prover~\cite{Wright94}.

In recent work~\cite{LMW10}, we have extended this encoding to object-oriented programs, by representing the memory of a state as a graph. Hence, a proof of refinement can be expressed as an Isabelle lemma, that needs to be proven by the user. Although we provide the user with a collection of lemmas to help her in this task, such a proof can still be challenging, in particular for users not familiar with Isabelle or theorem-proving in general. We address this challenge in this paper by providing a mechanism that automatically generates, when possible, the proof of the lemma in Isabelle.

This work takes place in the context of the refinement for Component and Object Systems (rCOS)~\cite{Xiceccs07,JSCP08}. The rCOS language has a formal semantics based on an extension of the Unifying Theories of Programming (UTP)~\cite{utp} to include the concepts of components and objects, and an operational semantics based on graphs~\cite{KLWZ09}.
This language is supported by the rCOS tool~\cite{LMS09}, which provides a UML-like multi-view and multi-notational modeling and design platform. The rCOS tool already provides the user
with a collection of complementary verification techniques, such as
the automated generation of robustness test cases~\cite{DBLP:journals/scp/LeiLLMS10}, and the automated generation of CSP processes to verify the compatibility between the sequence diagram  and the state diagram of a contract~\cite{CMS09}.

\paragraph{\bf Contribution} The main contribution of this paper is the extension of previous work on encoding the refinement of two rCOS programs as an Isabelle lemma~\cite{LMW10},
with a module that performs an automatic search for a sequence of pre-defined refinement steps showing that the refinement is correct. In general, a given program can be refined in different ways, and it is not always possible to predict the sequence of refinement steps, or even to know if it exists. If the module can find a correct sequence, then it directly generates the Isabelle proof of refinement for the initial lemma.
This module is written using the Maude rewriting tool~\cite{CDELMMQ99}, such that refinement rules are defined as rewriting rules, and each rewriting rule is associated with the corresponding Isabelle refinement lemma.

The main focus of this work is to present the architecture of the framework, thus describing how different environments (rCOS, Isabelle, Maude) interact together rather than presenting an encoding of a refinement calculus. Hence, we build upon
the previous encoding of the refinement calculus~\cite{Wright94,Laibinis00,Depasse01}, and only redefine what is necessary
for our integration with Maude.

\paragraph{\bf Organization}
The main novelty of this paper is the automatic generation of Isabelle proofs of refinement of rCOS programs using Maude.
In order to present and explain this automatic generation, we first briefly introduce the rCOS language together with
some simple illustrating examples in Section~\ref{sec:rcos}, we recall the previous mechanization of the
refinement calculus in Section~\ref{sec:calculus}, and we present our graph-based memory representation
of object-oriented programs together with the corresponding extension of the refinement calculus in Section~\ref{sec:graph}
and Section~\ref{sec:oorefinement}, respectively.

We then present the Maude module in Section~\ref{sec:proof_generation}, and provide an example
of generated proof in Section~\ref{sec:example}. We discuss the well-known aliasing problem
and show how to consider it in our approach in Section~\ref{sec:alias}.
Finally, we present related work in Section~\ref{sec:related_work} and conclude and present future work in Section~\ref{sec:conclusion}.

\section{rCOS}
\label{sec:rcos}

The rCOS method
consists of  two parts: a component/object-oriented language with  formal
semantics, and a modeling tool, enforcing a use-case based methodology for software development,
providing tool support and static analysis.  We use only a subset of the language in the examples presented in this paper, however we give here a brief description of the whole language, and we refer to~\cite{JSCP08} for further details.


The rCOS language is an extension of UTP~\cite{utp}, to include object-oriented and component features, and as such, the semantics of a program in any programming language can be defined as a predicate, called a {\em design}.  Roughly speaking, a design can be a traditional imperative statement, such as: an assignment \rcos{p := e}, where \rcos{p} is a path to a memory location and \rcos{e} is an expression; a conditional statement \rcos{$d_1$} $\lhd$ \rcos{b} $\rhd$ \rcos{$d_2$}, where \rcos{$d_1$} and \rcos{$d_2$} are designs and \rcos{b} is a boolean expression; a sequence \rcos{$d_1; d_2$}, where \rcos{$d_1$} and \rcos{$d_2$} are designs; a loop \rcos{$\pdo b\ d$}, where \rcos{$d$} is a design and \rcos{$b$} is a boolean expression; a local variable declaration and un-declaration \rcos{var T x = e} ; \rcos{end x}, where \rcos{T} is a type; an atomic design such as $\myskip$ ; $\chaos$.

A new object of type \rcos{$C$} is created and attached to the path \rcos{p} through the command
\rcos{$C$.new(p)}. A method invocation has the form \rcos{e.m(i; o)}, where \rcos{m} is a method, \rcos{i} stands for the input parameters and \rcos{o} for the output parameters. If there is no output parameter, we can write directly \rcos{e.m(i)}.

A design can also denote a more abstract specification, such as a pre/postcondition~\cite{Back78,Morgan94} \rcos{[pre(x) |- R(x,x')]}, meaning that if
the program executes from a state where the {\em initial value}
\rcos{x} satisfies \rcos{pre(x)}, the program will terminate in a state
where the final value  \rcos{x'} satisfies the relation \rcos{R(x,x')}
with the initial value \rcos{x}.
Similarly, non-deterministic choice is defined as \rcos{$d_1 \sqcap d_2$}, where \rcos{$d_1$} and \rcos{$d_2$} are designs.

The rCOS language includes the notion of components, which provide
or require contracts. A contract includes an interface (a set of field
and method declarations), the specification of each method and a
protocol stating the allowed sequences of method calls (for instance,
for a buffer, the method \rcos{put} must be called before the method
\rcos{get}). A component provides a contract through a class, where each method has to be defined
using a design. Note that the design of a method
can either be abstract (pre/post-conditions, non-deterministic choice, etc),
concrete (standard imperative and object-oriented features), or both at the same time, but
only concrete programs can be generated to Java.
For instance, all the following examples are correct rCOS programs.

\begin{minipage}{0.55\linewidth}
\begin{rcosenv}
class $A$ {
  int x;
  public m(int v) {
    x := v }
}


class $B_1$ {
  $A$ a;
  public foo() {
   [true |-a.x'=2 $\lor$ a.x'=3]}
}
\end{rcosenv}
\end{minipage}
\hfill
\begin{minipage}{0.45\linewidth}
\begin{rcosenv}
class $B_2$ {
  $A$ a;
  public foo() {
   [true |- a.x'=1] ;
   a.x:=a.x+1}
}
class $B_3$ {
  $A$ a;
  public foo() {
    a.m(1) ;
    a.x := a.x+1}
}
\end{rcosenv}
\end{minipage}

The method \rcos{$B_1$::foo} is abstract and non-deterministic: it just
specifies, under the true precondition, that the value of the field \rcos{x}
of the field \rcos{a} should be either equal to 2 or to 3.
The method \rcos{$B_2$::foo} mixes abstract pre/postconditions with
a concrete assignment while \rcos{$B_3$::foo} is completely concrete
and could be directly translated to Java. In this example, we can see
that \rcos{$B_1$::foo} is refined by \rcos{$B_2$::foo}, which is refined
by \rcos{$B_3$::foo}. We detail in the following section the mechanization
of the notion of refinement.

\section{Mechanized Refinement}
\label{sec:calculus}

The refinement calculus~\cite{Back78,Morgan94} is a program construction
method, where a non-deterministic specification is incrementally refined to
deterministic code, using pre-defined rules.
This calculus has been fully encoded into the theorem prover HOL, an ancestor of Isabelle, in ~\cite{Wright94,Depasse01} and then extended, in particular in~\cite{Laibinis00},
which introduces, among others, procedures and recursive functions.
The encoding follows the weakest precondition approach: for any design $d$ and any predicate $q$ over states, the function $\mathbf{wp}(d, q)$ stands for the weakest precondition that should be true on states before executing $d$ such that $q$ holds after executing $d$. Therefore, a design is usually considered as a predicate transformer, since it takes a predicate ($q$) as input and returns another predicate (the weakest precondition of $q$). We recall here the definitions of assignment and refinement from~\cite{Wright94}. We use \isa{State} to represent the type of
a program state, which is defined as a tuple of values in~\cite{Wright94}, and represented as a graph in~\cite{LMW10} and in this document. We introduce first the type of predicates over
states and the type of predicate transformers.

\begin{isaenv}
types  State pred =  State => bool
       State predT = State pred => State pred
\end{isaenv}

The \isa{assign} predicate transformer takes a function \isa{e},
which takes a state and returns the state where the corresponding
assignment is done. The weakest precondition of a predicate \isa{q}
is calculated by checking \isa{q} on a state where the assignment
has been done.

\begin{isaenv}
definition assign :: (State => State) => (State predT)
where
  assign e q == %u. q (e u)
\end{isaenv}

A design \isa{c1} is refined by a design \isa{c2} if, and only if, the weakest
precondition of \isa{c1} implies the one of \isa{c2} for any state.

\begin{isaenv}
definition implies :: (State pred) => (State pred) => bool
where
  implies p q == ALL u. (p u) --> (q u)

definition ref :: (State predT) => (State predT) => bool
where
  c1 ref c2 == ALL q. (implies (c1 q) (c2 q))
\end{isaenv}



Although the previous definitions do not directly
depend on the structure of the state, this structure is defined
as a tuple in~\cite{Wright94}, where each element of the tuple is the value of a variable
of the program. For instance, if a program has two variables
$x$ and $y$, set respectively to $1$ and $3$, the state of such a
program is the pair $(1, 3)$. The names of the variables are therefore
lost in the translation, and any operation concerning $x$ has to be
translated as an operation concerning the first element of the pair.
Dealing with local variables and method calls thus implies
to extend and narrow the state, respectively. 
Moreover, this approach does not directly handle references
and therefore such a representation for states cannot be applied
for object-oriented programs.
The usual way to tackle this issue
is to represent a state as a record or as a function from pointers
to values~\cite{tBaX03a,PO04,CN00,Sekerinski96}.
A recent approach uses graphs instead~\cite{KLWZ09}, and we present
it in the next section.

\section{Graph Representation}
\label{sec:graph}

In~\cite{KLWZ09}, the state of a program is represented as a directed labeled graph. We only give here a simple description of such a graph and its implementation in Isabelle/HOL, more details can be found in~\cite{KLWZ09,LMW10}.

\def\dol{\ensuremath{\$}}

\subsection{State Graph}
\begin{figure}
\centering
\subfigure[\label{fig:stategraph-a} State after execution of \rcos{v.a.b.x :=1; var int v=2}]{
  \begin{tikzpicture}[->,>=stealth',auto, node distance=1cm]
    \tikzstyle{every state}=[text=black]
    \tikzset{ n/.style = {circle, draw=black, font=\scriptsize}}
    \tikzset{ ed/.style = {font=\scriptsize}}
    \tikzset{ box/.style={ rectangle, rounded corners, draw=black,
       text width=1.5cm, minimum height=2em, text centered} }

  \node[n, label=above:{}] (p1) {$r_2$};
  \node[n, label=left:{}]  (p2)  [below left  of=p1]    {$2$};
  \node[n, label=right:{}] (r1)  [right of=p1, yshift=-0.0cm]    {$r_1$};
  \node[n, label=right:{}] (p3)  [below left of=r1, yshift=-0.0cm]    {$C$};
  \node[n, label=left:{}]  (p4)  [below left  of=p3]    {$B$};
  \node[n, label=right:{}] (p5)  [below right of=p3, yshift=-0.0cm]    {$A$};
  \node[n, label=right:{}] (p6)  [below left of=p5, yshift=-0.0cm]    {$1$};

  \path[<-] (p2) edge node [ed] {$v$} (p1);
  \path[->] (p1) edge node [ed] {$\$$} (r1);
  \path[->] (r1) edge node [ed] {$v$} (p3);
  \path[->] (p3) edge node [ed] {$a$} (p5);
  \path[->] (p5) edge node [above, ed] {$b$} (p4);
  \path[->] (p4) edge node [ed] {$x$} (p6);
  \end{tikzpicture}
  }
    \hspace{4cm}
\subfigure[\label{fig:stategraph-b} Recursive object]{
  \begin{tikzpicture}[->,>=stealth',auto, node distance=1cm]
     \tikzstyle{every state}=[text=black]
     \tikzset{ n/.style = {circle, draw=black, font=\scriptsize}}
     \tikzset{ ed/.style = {font=\scriptsize}}
     \tikzset{ box/.style={ rectangle, rounded corners, draw=black,
        text width=1.5cm, minimum height=2em, text centered} }

   \node[n, label=above:{}] (p1) {$r$};
   \node[n, label=right:{}] (p3)  [below right of=p1, yshift=-0.0cm]    {$C$};
   \node[n, label=left:{}]  (p4)  [below left  of=p3]    {$B$};
   \node[n, label=right:{}] (p5)  [below right of=p3, yshift=-0.0cm]    {$A$};
   \node[n, label=right:{}] (p6)  [below left of=p5, yshift=-0.0cm]    {$1$};

   \path[->] (p1) edge node [ed] {$v$} (p3);
   \path[->] (p3) edge node [ed] {$a$} (p5);
   \path[->] (p5) edge node [above, ed] {$b$} (p4);
   \path[->] (p4) edge node [ed] {$x$} (p6);
   \path[->] (p4) edge node [ed] {$c$} (p3);
   \end{tikzpicture}
  }
   \label{fig:stategraph}
   \caption{Examples of state graphs}
\end{figure}
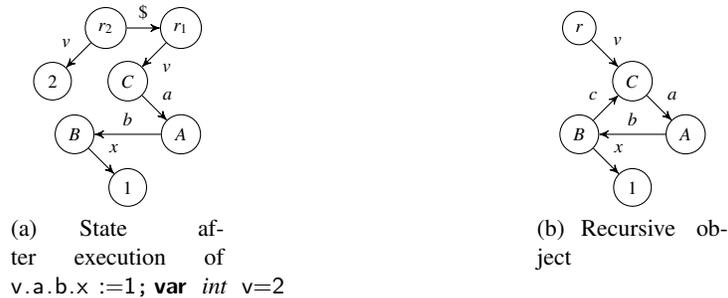

Intuitively, the state graph represents the abstract structure of the memory, such that an rCOS navigation path is represented
by a path in the state graph. Hence, a vertex can be either a root, a node or a leaf. A root represents a scope, and local
variables start from a root. A graph thus has a list of roots, one for each scope, and the root at the head of the list stands
for the current scope. For the sake of readability, we connect the roots using edges labeled by $\dol$ (the current root having
no incoming edge). For instance, Figure~\ref{fig:stategraph-a} illustrates the state of the graph after executing the statement
 \rcos{v.a.b.x :=1; var int v=2}. The creation of the new variable \rcos{v} leads to the creation of a new scope $r_2$,
from which this variable starts. When the scope exits, the node $r_2$ is removed, and so the newly created variable $v$ is no longer accessible; the previous root $r_1$ becomes current scope again.

A non-root vertex in a state graph represents either an object or a primitive datum,
called \emph{node} and \emph{leaf}, respectively.
A node is labeled by the runtime type of the object,
while a leaf is labeled by the primitive value.
An outgoing edge from a node is labeled by a field name of the source object and refers to
the target node or leaf representing the value of this field. There is no outgoing edge from a leaf.
Note that, as illustrated on Figure~\ref{fig:stategraph-b}, objects can be recursive.

 %


We implement the state graph in Isabelle by first introducing the datatype \isa{vertex}, which is the union of the base types
\isa{root}, \isa{node} and \isa{leaf}. We also consider a special vertex, $\bot$, which stands for the {\em undefined} vertex.
\begin{isaenv}
datatype vertex = N node | R root | L leaf  | $\bot$
\end{isaenv}

A state graph is defined as a list of roots, and a function taking a vertex $v_1$ and a label $l$ ({\em i.e.} a string for representing variable or field), and returns the vertex $v_2$ if $(v_1, l, v_2)$ is an edge of the graph, or $\bot$ otherwise. We write \isa{G} for the type of graphs.
\begin{isaenv}
types
 edgefun = vertex => label => vertex
 graph =  edgefun * root list
\end{isaenv}


In order to ensure that there is no edge starting with the
undefined vertex $\bot$, we introduce the following property\footnote{Note that we could equivalently constrain the range and domain of the function \isa{getEdgeFun},
however doing so tends to make the usage of this function more complex, since Isabelle does not provide
a native subtyping mechanism.}.
\begin{isaenv}
definition isGoodFunction:: graph => bool 
where
  isGoodFunction g == ALL x. (getEdgeFun g) $\bot$ x = $\bot$
\end{isaenv}
where \isa{getEdgeFun g} is used to get the first component of \isa{g}.

Moreover, we must ensure that the list of roots is consistent with the edge function, and so we say that a graph is well-formed, which we denote with the predicate \isa{wfGraph}, if, and only if, in addition to satisfy \isa{isGoodFunction}, each root is unique in the list of roots and is not the target of an edge, and there exists at least one outgoing edge for each root. We could similarly add the property that there are no outgoing edges from leaves, however, we do not need it in the current state of the development and therefore we chose not to add it. In general, it must be pointed out that we assume that the programs and memory states encoded in Isabelle are {\em generated} from correct rCOS programs. In other words, we do not expect the users to directly write programs in Isabelle, but instead to write them using the rCOS tool, which features a type-checker, and so can prevent by construction the generation of some not well-formed states, for instance one with an outgoing edge from a leaf. Hence, we only specify the state properties we need in order to prove the desired lemmas.


\subsection{Graph Operations}
This section briefly describes some basic graph operations that are needed for the encoding of the refinement calculus. Due to space limitation, we do not present here the implementation of these functions, more detailed explanations can be found in~\cite{LMW10}, and the complete list is available online\footnote{\sf http://www.doc.ic.ac.uk/$\sim$agriesma/mircos/graph\_utl.thy}.


We first introduce the type \isa{path} as a list of labels, which, for implementation optimization reasons, is reversed: the path \rcos{a.b.x} is represented by the list \isa{[''x'',''b'',''a'']}.
The vertex corresponding to a path \isa{p} in a graph \isa{g} is given by \isa{getVertexPath p g}.

The function \isa{swingPath} swings the last edge of a path in the graph to point
to a new vertex. In other words,
it sets a new value to a path in a state graph, and
therefore, can be used for implementing assignments in rCOS.
This function has been proved to preserve the well-formedness, \textit{i.e.}
for any graph \isa{g}, any path \isa{p} and any non-root vertex \isa{n},
if \isa{g} is well-formed then \isa{(swingPath p n g)} is also well-formed.
%

Intuitively, we would like to express the fact that after swinging a path to new vertex, this path actually points to this vertex. In practice, this is not true for every path. For instance, consider an infinite list such that \rcos{x.next} points to \rcos{x}. If we execute \rcos{x.next.next := y}, where \rcos{y} is another list, this is equivalent to executing \rcos{x.next := y}, and after the assignment, \rcos{x.next.next} points to \rcos{y.next}, which can be different from \rcos{y}.
This kind of situation happens when more than one prefixes of a path are aliasing with the owner of the path ({\em i.e.} not the vertex pointed by the path, but the one before).
Clearly, when dealing with such paths, some refinement rules do not hold any longer, for instance that  \rcos{[|- p' = e]} is refined
by \rcos{p := e} (the corresponding refinement lemma is given in Section~\ref{sec:ref_lemmas}).

Hence, we introduce the predicate \isa{isGoodPath}, such that given a path \isa{p}, \isa{isGoodPath p} is true if, and only if, there is only one prefix of \isa{p} that points
to the owner of \isa{p}.
For instance, in the state graph in Fig.~\ref{fig:stategraph-b}, the paths
\rcos{v.a}, \rcos{v.a.b.c} and \rcos{v.a.b.x} satisfy \isa{isGoodPath}, while
the paths \rcos{v.a.b.c.a} and \rcos{v.a.b.c.a.b.x} do not.
We can observe that the path \rcos{v.a}, which satisfies \isa{isGoodPath}, points to the same object as \rcos{v.a.b.c.a}, and we can safely replace  \rcos{v.a.b.c.a}
by  \rcos{v.a}, and obtain an equivalent program. By extending this observation to all paths, 
we can assume that any path that does not satisfy \isa{isGoodPath}
can be replaced by a path that does, following the idea that any cycle can be removed from a path in a graph. 
We assume here that this substitution is
done at the rCOS level, and that any path considered in Isabelle satisfies the predicate \isa{isGoodPath}.

Furthermore, a path \isa{p} is said to be well-formed with respect to \isa{g}, denoted by \isa{wfPath p g},
if, and only if,  the vertex of \isa{p} exists in \isa{g}, and \isa{p} satisfies \isa{isGoodPath}.
Note that since the paths are generated from a correct rCOS program, and since rCOS has a type-checker,
it follows that the paths are well-typed, and therefore that the vertex pointed by a path always exists in a graph.



One of the most important theorems of our theory is \isa{swingPathChangeVertex}: given a well-formed graph,
swinging a well-formed path to a new vertex makes this path point to this vertex in the resulting graph.



The function \isa{Vars} combines the operations of creating a root vertex and adding edges, and therefore
implements local variable declaration.
In a similar way, the function \isa{removeSnode}
removes the top root from the root list, and in consequence
all edges outgoing from the root. It implements local variable un-declaration.
Finally, the function \isa{addObject} creates a new node vertex (object)
in a graph. These functions are proved to preserve the graph well-formedness.

%

\section{Refinement of rCOS Designs}
\label{sec:oorefinement}

The graph-based representation of the memory presented in the previous
section allows us to extend the mechanization of the refinement calculus
presented in Section~\ref{sec:calculus} to deal with object-orientation.
Since we only consider well-formed graphs and paths,
we integrate these conditions into the weakest precondition of each command.
The complete definition of the refinement calculus for all constructs is available online\footnote{\sf http://www.doc.ic.ac.uk/$\sim$agriesma/mircos/rcos.thy}.

\subsection{Primitive Designs}

\subsubsection*{Pre/post-condition}
The definition of the non-deterministic assignment, which stands for a predicate between the next graph and the past one,  needs to include
the well-formedness checks.
\begin{isaenv}
definition nondass :: (graph => graph pred) => path list => (graph pred) => (graph pred)
where
nondass P l q == (%v. (wfGraph v) & (wfPathl l v) & (ALL v1. P v v1 --> q v1))
\end{isaenv}
where \isa{wfPathl l v} is true if, and only if, every path in \isa{l} satisfies \isa{wfPath}.
This list of paths corresponds to all the paths appearing in the postcondition.
A pre/postcondition is then an assertion followed by a non-deterministic
assignment.
\begin{isaenv}
definition pp :: (graph pred) => (graph => graph pred) => path list => (graph predT)
where
  pp p r l == assert p ; nondass r l
\end{isaenv}
where \isa{assert} is the standard definition for the assertion.
For instance, the design of the method \rcos{$B_1$::foo} given in Section~\ref{sec:rcos} and defined by \rcos{[ true |- a.x'=2$\lor$a.x'=3 ]} is translated into the statement:
\begin{isaenv}
pp (true) (% g. % g1. ((getNVal $this.a.x$ g1) = 2 | (getNVal $this.a.x$ g1) = 3)) [$this.a.x$]
\end{isaenv}
where \isa{getNVal} is the function returning the value (as a \isa{nat}) of the path  in the given graph and where, for the sake of readability, we abbreviate the path \isa{[''x'',''a'',''this'']} as $this.a.x$.

\subsubsection*{Assignment}

The definition of the assignment is changed as follows.
\begin{isaenv}
definition assign :: path => exp => (graph pred) => (graph pred)
where
  assign p e q == %u. wfGraph u & wfPath p u
                  & wfExp e u & q (swingPath p (getNodeExp e u) u)
\end{isaenv}
where the path \isa{p} is assigned to the expression
\isa{e}, which is required to
be well-formed.
The function \isa{getNodeExp}
returns the value of an expression, which is obtained using
\isa{getVertexPath} when the expression to be evaluated is a path,
otherwise itself when it is a constant value. For instance, the
assignment \rcos{a.x := a.x+1} of the method \rcos{$B_2$::foo} given in Section~\ref{sec:rcos}
is translated as:
\begin{isaenv}
assign $this.a.x$ (Plus (Path $this.a.x$) (Val (Zint 1)))
\end{isaenv}

\subsubsection*{Local declaration and un-declaration}
The commands \isa{begin} and \isa{end}
declare/initialize new local variables and  terminate them, respectively.

\begin{isaenv}
definition begin :: labelExpF => (graph pred) => (graph pred)
where
  begin f q == %u. wfGraph u & wflabelExpF f u & q (Vars f u)

definition end :: (graph pred) => (graph pred) 
where
  end q == %u. wfGraph u & q (removeSnode u)
\end{isaenv}
where \isa{f} is a well-formed function of type \isa{labelExpF}
(i.e. \isa{label => exp}, mapping variables to their initial expressions),
which means that for each local variable,
it is initialized by a well-formed expression
in \isa{f}.

The command \isa{locdec} defines the block for local declaration and un-declaration, where \isa{f}
is the same as above and \isa{c} is the body of the block.
\begin{isaenv}
definition locdec :: labelExpF => (graph predT) => (graph predT)
where
  locdec f c == begin f; c; end
\end{isaenv}

\subsubsection*{Method invocation}

The command \isa{method} implements a method invocation with the help of
the command \isa{locdec}.
\begin{isaenv}
definition method :: (label * exp) list => (graph predT) => (graph predT)
where
  method l c == locdec (getLabelExpF l) c
\end{isaenv}
where \isa{l} is of type \isa{(label * exp) list}, each pair consisting of
a formal parameter and its actual value, and \isa{c} is the method body followed
by the assignment from the formal return parameter to the actual
return parameter. In the \isa{method} command, the function \isa{getLabelExpF}
translates a list of pairs of type \isa{label * exp}
to the corresponding mapping of type \isa{labelExpF} (i.e. \isa{label => exp}).
For instance,
the method call \rcos{a.m(1)} of the method \rcos{$B_3$::foo} of Section~\ref{sec:rcos}
is translated as:
\begin{isaenv}
method [($this$, Path $this.a$), ($v$, Val (Zint 1))] ; assign $this.x$ Path $v$
\end{isaenv}
When the method is called, the variable $this$ is substituted by
$this.a$ (the caller), and $v$ by $1$.
Note that with this approach, recursive method calls
are not directly handled, and require the definition of a fix-point,
which we do not consider here.




\subsection{Composite Designs}

With the predicate transformer semantics, the definitions of the
composite designs, like the sequential composition, the loop or the
conditional statement, do not depend on the representation of the
memory state. Hence, we can directly re-use the definitions and theorems
from~\cite{Wright94}.  For instance, the sequential composition
\isa{c; d} is refined by \isa{e; f} if \isa{c} is refined by \isa{e}
and \isa{d} is refined by \isa{f} and \isa{c} is monotonic, and in
fact, we have proved that all basic commands (i.e. \isa{nondass},
\isa{pp}, \isa{assign}, \isa{begin} and \isa{end}) are monotonic, and
the compound constructs \isa{locdec, method, cond, do, seq} preserve
monotonicity with respect to their subcomponents.  Moreover, the other
constructs such as the conditional \isa{cond} and the loop \isa{do}
preserve refinement with respect to their subcomponents.  By applying
these theorems, we can refine a program by repeatedly refining its
subcomponents, and then prove that the new generated program is a
refinement of the old one.

\subsection{Tool Refinement}
\label{sec:tool_refinement}

Refining a model is, by definition, a dynamic process: a new model is generated from a previous one, by applying some refinement rules. The main challenge is then to be able to consider both models at the same time, in order to generate the corresponding proof obligations.  When the refinement concerns only method bodies, the rCOS tool provides a simple way to define a refinement operation.  Firstly, a class is created, and stereotyped with a specific kind of refinement, for instance refining automatically every \rcos{[true |- x'=e]} by \rcos{x := e}. The most general refinement is the manual refinement, where the user provides an operation, its old design (mainly for sanity checks), and the refining design.  In a second step, the user can, at any time, apply such a refinement by right-clicking on the corresponding class and selecting the ``refine'' operation, and the tool then transforms the model accordingly.

\bigskip

For instance, the user can indicate can create a new class to specify that the design of \rcos{$B_1$::foo} is refined by the
design of \rcos{$B_3$::foo}, given in Section~\ref{sec:rcos}, {\em i.e.} the user
wants to prove that the design \rcos{[ true |- a.x'=2$\lor$a.x'=3 ]} is refined
by \rcos{ a.m(1) ; a.x := a.x + 1}. 
In this case, the rCOS tool generates the following Isabelle lemma:
\begin{isaenv}
lemma b1_foo_ref_b3_foo :
  "pp (true) (% g. % g1. ((getNVal $this.a.x$ g1) = 2 | (getNVal $this.a.x$ g1) = 3)) [$this.a.x$]
  ref
  ((method [(''this'', Path $this.a$), (''v'', Val (Zint 1))] (assign $this.x$ (Path [''v'']))) ;
    assign $this.a.x$ (Plus (Path $this.a.x$) (Val (Zint 1))))"
\end{isaenv}
Note that at this stage, the rCOS tool only generates the statement of the lemma, and not the proof. The user
can either try to prove it manually, or to use our Maude module, presented in the following section.

\section{Automatic Proof Generation}
\label{sec:proof_generation}

\begin{figure}[tp]
\begin{center}
\begin{tikzpicture}[->,>=stealth',auto, node distance=2cm]
\tikzstyle{every state}=[text=black]
\tikzset{ n/.style = {circle, fill=black}}
\tikzset{ box/.style={ rectangle, rounded corners, draw=black,
       text width=1.5cm, minimum height=2em, text centered} }

\node[box] (rcos)     {rCOS tool};

\node (p1) [above of=rcos, xshift=-0.5cm, yshift=-1cm] {$p_1$};
\node (p2) [above of=rcos, xshift=0.5cm, yshift=-1cm] {$p_2$};

\node[box]  (maude) [below of=rcos] {Maude};

\path[->] (p1) edge (rcos);
\path[->] (p2) edge (rcos);
\path[->] (rcos) edge node [left] {$\{p_1 \sim > p_2\}$} (maude);

\node[box] (pe) [right of=maude, xshift=2.5cm]  {Proof Extracteur};
\path[->] (maude) edge node [below] {$p_1 \stackrel{r_0}{\rightarrow} q_1 \cdots q_n \stackrel{r_n}{\rightarrow} p_2 $} (pe);

\node[box] (isa) [above of=pe] {Isabelle};

\path[->] (rcos) edge node  {{\tt lemma $p_1$ ref $p_2$}} (isa);
\path[->] (pe) edge node [left] {
    \begin{tabular}{c}
        {\tt $p_1$ ref $q_1$ by $r_0$} \\
        $\cdots$\\
        {\tt $q_n$ ref $p_2$ by $r_n$}
    \end{tabular}} (isa);
\end{tikzpicture}
\caption{Workflow \label{fig:workflow}}
\end{center}
\end{figure}
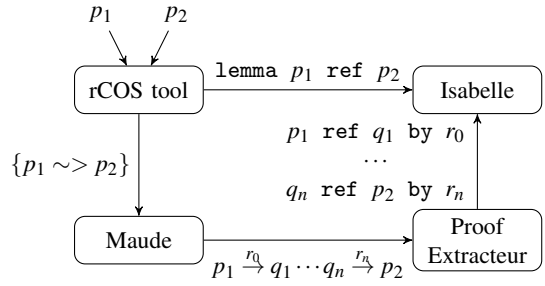

Intuitively, we define for each refinement step a rewriting rule and an Isabelle lemma,
such that when the rewriting rule is applied, the corresponding refinement step can be
directly proven. Hence, the global architecture of our system, illustrated in Fig.~\ref{fig:workflow}, can be described as follows: given two rCOS programs $p_1$ and $p_2$, the rCOS tool generates both the Isabelle statement \isa{lemma $\,\,p_1\,\,$ ref $\,\,p_2\,\,$}, as described in Section~\ref{sec:tool_refinement}, and the Maude term \maude{$(id\,\,\{ p_1\} \sim> \{p_2 \}\,\, \mathit{status}: s$)}, that we describe in the following.
When the Maude module can find a rewriting sequence, then it generates the Isabelle proof for the lemma.

We first introduce some examples of refinement lemmas we provide, then we briefly introduce Maude~\cite{CDELMMQ99} and present the rewriting rules corresponding to the refinement rules, and finally show how to extract an Isabelle proof. A detailed example is given in Section~\ref{sec:example}.

\subsection{Refinement Lemmas}
\label{sec:ref_lemmas}

In addition to the theorems introduced in~\cite{Wright94}, we provide
lemmas corresponding to refinement steps. For the sake of simplicity, we only focus here on lemmas concerning integers, however equivalent lemmas can be defined for other primitive types.

For instance, the lemma stating that for any path \rcos{p} and any integer \rcos{n}, the statement
\rcos{[true |- p'=n]} is refined by the assignment \rcos{p := n} is
defined\footnote{Due to space limitation, we do not include in this document the proofs of the lemmas, which can be found at
{\sf \footnotesize http://www.doc.ic.ac.uk/$\sim$agriesma/mircos/rcos\_lib.thy}.}
 as:
\begin{isaenv}
lemma ref_pp_assign :
  "pp (true)  (% g. % g1.((getNVal p g1) = n)) [p] 
   ref
   (assign p (Val (Zint n)))"
\end{isaenv}
Another example is the Expert Pattern, which is an essential rule for object-oriented
functionality decomposition by delegating responsibilities
through method calls to the objects, called the experts, that have the
information to carry out the responsibilities.
For instance, defining a setter for a field is a special case of the
Expert Pattern, and therefore a refinement.
As a special instance of this pattern, we have defined the lemma \isa{EPIsRefTwo},
which states that the statement \rcos{p.a := n} is refined by the method
\rcos{p.m(n)} where \rcos{m (T v) \{this.a := v\}} is a method of
\rcos{p}, for any primitive type \isa{T} and parameter \isa{v}.
\begin{isaenv}
lemma EPIsRefTwo :
"p~=[] --> b ~= c --> 
 (assign (p.a) (Val n) 
 ref
  (method [(b, Path p), (c, Val n)] (assign (b.a) (Path [c]))))"
\end{isaenv}
This lemma only considers attribute accesses (if \isa{p} is empty, then \isa{p.a} represents a local variable and
Expert Pattern is not necessary), and that \isa{b} and \isa{c} are the formal parameters of the method, and therefore must be different.

Finally, the lemma \isa{assign_end} states that given a path \isa{p} and two integers \isa{m} and \isa{n}, the statement \isa{p := m} is refined by \isa{p := n; p := p + (m - n)}.
\begin{isaenv}
lemma assign_end :
"(assign p (Val (Zint m))) ref
  (assign p (Val (Zint n)); (assign p (Plus (Path p) (Val (Zint (m - n))))))"
\end{isaenv}

\subsection{Maude} 
\label{sub:maude}

Maude is a rewrite tool that allows the specification of
\textit{equations} and \textit{rewrite rules} which have a simple
rewriting semantics in which instances of the left hand side are
replaced by corresponding instances of the right hand side.  The set
of equations is designed to be confluent and terminating.  This means
that starting from a term, every possible sequence of applications of
equations leads to a canonical form made from \emph{ground terms},
which also give the means to define the type system of the implemented
logic.  Application of rewriting rules, on the other hand, needs
neither be confluent nor terminating, and allows to express the
evolution of the system.  Eligible rules are selected by pattern
matching: while \textit{equations} have a deterministic result for any
enabling term and are executed immediately by the Maude engine, the
order of execution of \textit{rewrite rules} may lead to different
results.  Thus, the application of rewrite rules spans a state space
that can be explored by choosing among enabled rewrite rules.
Intuitively, we thus define a state as a set of proof obligations that
is known at any point in the execution and will use rewrite rules to
generate new proof obligations and search for the proof, while the
equations provide us with a canonical representation of equivalent
terms, and discharge simple proof obligations.

\subsection{Rewriting Rules} 
\label{sub:rewriting_rules}

In order to search for a sequence of steps to form a proof of
refinement, we use Maude to systematically perform syntactical
rewriting starting with an initial proof obligation that represents
the refinement of a specification by an implementation.
In the context of Maude, we write a proof obligation as \maude{$(id,
  \{p_1\} \sim> \{p_2\},\; \mathit{status:} s\; \mathit{from:} f)$},
where $id$ is an identifier of the proof-obligation, \maude{$\{p_1\}
  \sim> \{p_2\}$} stands for the refinement step of $p_1$ to $p_2$,
$f$ refers to further proof-obligations that need to be fulfilled such
that the rule can be discharged, and $s$ is the status of the
proof-obligation.  The status can either indicate that it is still
undetermined how to discharge the proof-obligation, (marked as
\maude{todo}), or it gives a lemma with which it should be discharged
in Isabelle.

We distinguish between two kinds of rewrite actions: 1)~setting up new
proof obligations according to the syntax of the present obligations
and 2)~discharging proof obligations according to the the basic
refinement steps proved by Isabelle as described in the previous
sections.  These kinds of rules often come in pairs where the former
type of rules generates the proof obligations that are required by the
latter, as shown below for \maude{ref-sequential}.  Only when all
dependencies are proven, their \maude{id} is stored in the
$\mathit{from}$ field and the proof obligation is discharged.  Note
that some of those steps are implemented as equations for performance
reasons.  This can be done if the application of the rule is
definitely required and the outcome is deterministic.  E.g., the
refinement between two identical terms is immediately discharged by
the \maude{ref-reflexive} rule.  Other rules introduce new
proof-obligations on a speculative basis.  Such rules may or may not
be required in the process of finding a proof and therefore, according
to the definitions of equations and rewrite rules given above, are
implemented as rewrite rules.  For simplicity of presentation, we give
here all of the steps in form of rewrite rules.

In the following, we present some selected rules for generation and
discharging of proof obligations as rewrite rules\footnote{The
  complete definition in maude rewriting logic can be found at {\tt
    http://www.doc.ic.ac.uk/\~{}agriesma/mircos/rcos.maude}.}.  The
syntax for expressions and statements in Maude is very similar to the
syntax of the Isabelle lemmas; in fact, the communication between the
tools can be done by a simple maude export expression and a script
that replaces some reserved key symbols, in the remainder of the
section we stick to a slightly simplified presentation of the rules;
in particular, we omit the environment taking track of details like
number of open obligations and similar.  We use \maude{S1..S4} to
denote statements, \maude{E1, E2} for expressions and \maude{X} for
variables.
%
%
The presented rules (\texttt{rl}) have the following structure:
\begin{maudeenv}
rl [$name$] $po_1, \cdots, po_k$ => $po_{k+1},\cdots, po_n$ .
\end{maudeenv}
where $name$ identifies the rule, and $po_i$ are
proof-obligations.  At least one of $po_1 \dots po_k$ is undischarged,
marked by a status equal to \maude{todo}.  Conditional rewrite rules
(\texttt{crl}) have an additional ``\maude{if E1}'' clause that needs
to evaluate to true for the rule to be enabled.  For instance, the
rule \maude{ref-reflexive} can be defined as:
\begin{maudeenv}
rl [ref-reflexive] :
  ( $id$ { X } ~>   { X } status:todo ) =>
  ( $id$ { X } ~>   { X } status:ref-reflexive ) .
\end{maudeenv}
This rule can be read as follows: if we need to discharge the
proof-obligation that the program \maude{X} refines the program
\maude{X}, then we can directly do so by using the Isabelle lemma
\isa{ref-reflexive}.  New proof obligations are introduced by pattern
matching of present proof obligations.  For instance, the refinement
rule \maude{ref-sequential-gen1} matches for sequences of statements
and introduces a new proof obligation (note that $id_2$ is a fresh
identifier).
\begin{maudeenv}
crl [ref-sequential-gen1] :
  ( $id_1$ { S1 ; S2 } ~>  { S3 ; S4 } status: todo ) =>
  ( $id_1$ { S1 ; S2 } ~>  { S3 ; S4 } status: todo ) ,
  ( $id_2$ { S1 } ~>  { S3 }           status: todo )
if new( S1, S3 ) .
\end{maudeenv}
This rule represents the fact that in order for the left hand side
(LHS) to be refined by the right hand side (RHS), the first statement
on the LHS, \maude{S1}, needs to be refined by the first statement on
the RHS, \maude{S3}. (An analogous rule exists for the second
statement.)  The rule is conditional and only creates a new proof
obligation if \maude{$\{S_1\} \sim> \{S_3\}$} is not present yet.
Similar rules exist for other constructs and can also match multiple
present proof obligations to, e.g., create a missing part for the
lemma of transitivity.

The corresponding discharge rule for refinement of sequences makes
sure that the proof that \maude{S1 ; S2} is refined by \maude{S3 ; S4}
is only discharged if, and only if, we have discharged
proof-obligations $id_2$, stating that \maude{S1} is refined by
\maude{S3}, and $id_3$, stating that \maude{S2} is refined by
\maude{S4}.  This is recorded by setting the \textit{status:} to the
Isabelle lemma \isa{ref-sequential}, with its arguments $id_2$ and
$id_3$ stored in the \textit{from:} field:
\begin{maudeenv}
crl [ref-sequential] :
 ($id_1$ { S1;S2 } ~> { S3;S4 } status: todo )
 ($id_2$ { S1 } ~> { S3 }           status: stat1 ) ,
 ($id_3$ { S2 } ~> { S4 }           status: stat2 ) ,
=>
 ($id_1$ { S1;S2 } ~> { S3;S4 } status: ref-sequential from: $id_2$ $id_3$)
 ($id_2$ { S1 } ~> { S3 }           status: stat1 ) ,
 ($id_3$ { S2 } ~> { S4 }           status: stat2 )
if stat1 != todo and stat2 != todo .
\end{maudeenv}

The provided rules may not always be sufficient to fully discharge all
proof obligations.  E.g., the rule \maude{ref-strengthen} refines a
pre/postcondition by replacing the postcondition by another one that
logically implies it as follows:
\begin{maudeenv}
rl [ref-strengthen] :
 ($id$ {[ |- E2 ]} ~> {[ |- E1 ]} status: todo ) =>
 ($id$ {[ |- E2 ]} ~> {[ |- E1 ]} status: ref-strengthen from: $id_2$),
 ($id_2$ prove (E1 $\Rightarrow$ E2)        status: sorry ) .
\end{maudeenv}
We use the Isabelle keyword \maude{sorry} to denote that the
proof-obligation {\em cannot} be discharged in Maude, and therefore
has to be done in Isabelle, where this keyword allows one not to provide
the proof of a lemma. This approach makes it possible to consider the
postcondition strengthening refinement rule regardless of the
postcondition itself, by delegating the burden of the proof to
Isabelle. However, some instances of this rule are quite simple, and
can be done directly in Maude. For instance, the rule \maude{ref-disj-left}
chooses a member of the disjunction in a postcondition.
\begin{maudeenv}
rl [ref-disj] :
 ($id$ {[ |- E1 \/ E2 ]} ~> {[ |- E1 ]} status: todo ) =>
 ($id$ {[ |- E1 \/ E2 ]} ~> {[ |- E1 ]} status: ref-disj-left) .
\end{maudeenv}

A major strength of this approach is its extensibility: new refinement rules can be easily added, simply by adding the corresponding rewriting rules in the Maude file, and the corresponding Isabelle lemmas in the Isabelle file, without any required modification to existing code. Hence, sets of rules can be dynamically loaded and unloaded, to adapt to different contexts.



\section{Example} 
\label{sec:example}
We consider the lemma \isa{b1_foo_ref_b3_foo} from
Section~\ref{sec:tool_refinement}, which expresses that the design of
the method \rcos{$B_1$::foo} is refined by the design of
\rcos{$B_3$::foo}, given in Section~\ref{sec:rcos}. The following
maude term is generated by the Maude module:
\begin{maudeenv}
{[ |- 2 = $a.x$ ' $\vee$ 3 = $a.x$ ']} ~>
{method[(["this"],["a"]),["v"],1]($this.x$ := ["v"]) ; $a.x$ := 1 + $a.x$} status: todo])
\end{maudeenv}
where, here again, we abbreviate paths.  Maude applies enabled rules
until a proof for the refinement is found.  The sequence of rewrite
rules applied to generate the proof below is shown in
Table~\ref{tab:minimal}.  We see the rules that correspond to actual
lemmas in the Isabelle proof (e.g., \maude{ref-transitive})
interleaved with rewrite rules that tentatively generate new proof
obligations for the search (rules containing \maude{gen}).
\begin{table}
\centering
\begin{minipage}{.48\linewidth}
\begin{tabular}{l}
ref-mcall-gen \\
ref-sequential-gen1\\
ref-sequential-gen2\\
is-ident\\
ref-mcall\\
ref-sequential\\
ref-transitive-gen-left\\
ref-add-gen\\
ref-add\\
ref-transitive-gen-left\\
ref-pp-assign\\
ref-disj-left\\
ref-transitive-gen-right\\
ref-pp-assign\\
ref-transitive\\
ref-transitive\\
ref-transitive\\
\\
\\
\end{tabular}
\caption{minimal rewrite steps}\label{tab:minimal}
\end{minipage}
\begin{minipage}{.48\linewidth}
\begin{tabular}{l}
ref-mcall-gen\\
ref-sequential-gen2\\
ref-sequential-gen1\\
is-ident\\
ref-mcall\\
ref-sequential\\
ref-transitive-gen-left\\
ref-add-gen\\
ref-add\\
ref-transitive-gen-left\\
ref-pp-assign\\
ref-disj-left\\
ref-transitive-gen-right\\
ref-transitive-gen-right\\
ref-pp-assign\\
ref-transitive\\
ref-transitive\\
ref-transitive\\
ref-transitive
\end{tabular}
\caption{alternative sequence}\label{tab:alternative}
\end{minipage}
\end{table}
By default we select the shortest proof.  Note however, that this is
not a unique solution --- an alternative, slightly longer sequence of
steps is shown in Table~\ref{tab:alternative}.  We see that,
additionally to the order in which new obligation are created, we also
may have additional generation- and discharging rules (e.g., the
alternative proof contains four \maude{ref-transitive} lemmas).
Searching for the shortest proof is not trivial and can be abandoned
in favor of a longer path if a predefined time limit is exceeded.  The
found proof for Table~\ref{tab:minimal} is given in the following (for
the clarity of the presentation, we have manually added the definition
of the \isa{?X} terms, the proof being equivalent without them, but
much harder to read):
\begin{isaenv}
proof -
  let ?A = "(pp (% g. True) (% g. % g1.((getNVal $this.a.x$ g1) = 2 )) [$this.a.x$] )"
  let ?B  = "(assign $this.a.x$ (Val (Zint 2)))"
  have f9: "?A ref ?B" by (simp add:ref_pp_assign)
  let ?C = "pp (% g. True) (% g. % g1.((getNVal $this.a.x$ g1) = 2
                           | (getNVal $this.a.x$ g1) = 3)) [$this.a.x$ ]"
  have f8: "?C ref ?A"    by (simp add:ref_disj_left)
  from f8 f9  have f7: "?C ref ?B" by (simp add:ref_transitive [of ?C ?A ?B])
  let ?D = "(assign $this.a.x$  (Val (Zint 1))) ; (assign $this.a.x$  (Plus (Path $this.a.x$) (Val (Zint 1))))"
  have f6: "?B ref ?D"
    by (insert assign_end  [of $this.a.x$  2 1], simp )
  from f7 f6  have f5: "?C ref ?D" by (simp add:ref_transitive [of ?C ?B ?D])
  let ?E = "assign $this.a.x$ (Val (Zint 1))"
  have f4: "monotonic ?E" by (simp add:assign_monotonic)
  let ?F = "(method [(''this'', (Path $this.a$)), (''v'', (Val (Zint 1)))]
               (assign $this.x$ (Path [''v''])))"
  let ?G = "assign $this.a.x$ (Plus (Path $this.a.x$) (Val (Zint 1)))"
  have f3: "?G ref ?G"  by (simp add:ref_reflexive)
  have f2: "?E ref ?F"  by (simp add:EPIsRefTwo)
  from f2 f3 f4  have f1: "?D ref ?F ; ?G" by (simp add:seq_ref)
  from f5 f1 have f0: "?C ref ?F ; ?G"
    by (simp add:ref_transitive [of ?C ?D "?F ; ?G"])
  from f0 show ?thesis by simp
qed
\end{isaenv}

This proof is correct, and can be instantly verified by Isabelle. Note that although it is quite long for a simple lemma, each step is atomic, since only one lemma is used for each step. In practice, it is possible to come up with a shorter, but equivalent proof of this lemma, by manually "inlining" the facts: every time that a fact $f_i$ is used to prove a fact $f_j$, we try to prove directly $f_j$ by adding the tactics used for $f_i$. If it succeeds, then $f_i$ can be removed. However, in general, it is not trivial to find out which facts can be removed.

\section{Aliasing}
\label{sec:alias}

In an object-oriented program, an accessible object may be referred to by multiple navigation paths, which are aliasing to each other. Because an object can be modified via any alias, the behavior of object-oriented programs is hard to specify and verify. Recently, object aliasing has been extensively studied, and many methods and techniques for aliasing analysis and control have been proposed, in particular shape analysis~\cite{Sagiv:1999}, separation logic~\cite{Reynolds02separationlogic:}, and ownership types~\cite{Clarke:1998}.  By applying these different techniques, it can be checked whether two expressions in a program execution may become aliased, and furthermore, provided with the known alias relations, what properties a program will satisfy.

Our graph-based implementation of program states makes object aliasing easy to interpret: two paths are \emph{aliasing} in a graph if, and only if, they refer to the same vertex in the given state graph. We therefore introduce the predicate \isa{alias},
and we can easily prove the following lemma\footnote{The proof of this lemma, together with the example presented below, can
be found at {\sf \footnotesize http://www.doc.ic.ac.uk/$\sim$agriesma/mircos/alias.thy}.}:
\begin{isaenv}
lemma aliasPreservesAssertAssign:
" implies q (alias p1 p2) ==> implies q  (wfPath (p2.a)) ==>
 assert q ; assign (p1.a) x ref (assert q; assign (p2.a) x)"
\end{isaenv}
This lemma states that if \rcos{p1} and \rcos{p2} are aliases, then \rcos{p1.a := x} is refined by \rcos{p2.a := x}.
The hypothesis with \isa{(wfPath (p2.a))} can be in practice automatically discharged, since the generation from rCOS to Isabelle ensures to consider only well-formed paths. Note that
we provide the rCOS developer with a built-in predicate \rcos{alias(p1, p2)}, allowing her to express that two paths are
aliasing. It becomes then possible to prove that \rcos{[alias(a, b) |- a.x' = 3]} is refined by \rcos{b.x := 3}.
The appropriate rewriting rules and lemmas for aliasing are included in our Maude module, and this refinement can be proven automatically.


Although reasoning with aliases is directly possible in our framework, calculating the aliasing relationship proves to be much
harder. Indeed, as we never generate the actual graph, but only graph transformation operations, we cannot directly check if two paths are aliasing. Moreover, it is not trivial to statically find out if two paths are aliasing. For instance, an intuitive
rule could say that when assigning the value of a path \rcos{p1} to a path \rcos{p2}, \rcos{p1} and \rcos{p2} become aliases. However, this statement clearly does not hold, for instance if \rcos{x} represents a traditional linked list, after the
assignment \rcos{x.next := x.next.next}, \rcos{x.next} and \rcos{x.next.next}, if they are different from \rcos{null}, are
not aliasing. Hence, at this point of our development, we rely on the user to provide the aliasing statements, through
assertions and/or preconditions, and we assume that these statements can be proved by another tool, such as those mentioned at
the beginning of this section.



\section{Related Work - Discussion} 
\label{sec:related_work}

\subsection{Mechanization of Refinement} 
\label{sub:mechanization_of_refinement}

The mechanization of the refinement calculus was firstly done
in~\cite{Wright94}, which has been extended to include
pointers~\cite{tBaX03a} and also object-oriented
programs~\cite{CN00,Sekerinski96}. In particular, a refinement calculus has been defined for Eiffel contracts~\cite{PO04}, and encoded in PVS~\cite{Paige03formalisingeiffel}. Although this approach addresses a similar issue than the one exposed here, the authors encode the calculus using a shallow embedding, that is, a class in Eiffel is encoded as a type in PVS, a routine in Eiffel is encoded as a function in PVS, etc. Proofs of refinement are then done over PVS {\em programs} rather than PVS {\em terms}, and so require the understanding of the underlying semantics of PVS. We use here a deep embedding, following~\cite{Wright94}, and the proofs of refinement are done, roughly speaking, over the abstract syntax tree of the original program, and so only require to know how to write a proof in Isabelle/Isar. The Program Refinement Tool~\cite{Carrington96atool} provides a deep embedding of a refinement calculus, and even if it does not support \textsc{OO} programs natively, it could be extended with an existing formalization which does~\cite{Utting92anobject-oriented}. However, rCOS also provides a semantics for components, and even if we do not address in this paper the issue of verification of component protocols, this work is part of a larger framework where other verification techniques exist~\cite{StolzUMLFM09}. In other words, the work presented here is not a standalone tool, but adds up to a collection of tools that helps a
developer to specify, implement and verify an application.

\subsection{Certified Model Transformations} 
\label{sub:model_transformations}

The next step is therefore to express the refinement of models rather of simple statements, in particular for
model transformations, which is an on-going work
in the rCOS tool~\cite{StolzUMLFM09}. The principal challenge
in this work is for the tool to handle several models at the same time: before, during and after refinement.
For instance, in the example we have presented, we assume that the method \rcos{m} is already present in the class \rcos{A}. However, in practice, the software engineer might want to create the setter and change the code at the same time. In this case, creating the setter is a correct refinement, but in order to prove it, we need to also encode the structure of the whole model in Isabelle, in order to express that a whole model refines another one, and then define the model transformations in Isabelle. A related approach using the Coq theorem prover has been recently proposed~\cite{DBLP:conf/sbmf/CalegariLST10}, where the Class to Relational model transformation has been certified.

\subsection{Memory Model} 
\label{sub:memory_model}

Different memory
models for object-oriented programs have been encoded in theorem
provers~\cite{Filliatre03,Berg01theloop,TNipkow}. However, the
memory in these approaches is either modeled as a function from
addresses or pointers to values or using records to represent
objects. Although such a modeling is very expressive, and has
been shown to be adapted to automated demonstration, we propose
here a  representation of the memory by a directed and labeled
graph, that might be more visual than a representation by a function or set of records.
The graph structure helps in the formulation of properties and
carrying out interactive proofs.


\subsection{Automation of Refinement} 
\label{sub:automation_of_refinement}

A range of tools are developed for supporting automatic refinement.
Inspired by Dijkstra's
weakest precondition calculus, they usually
generate proof obligations corresponding to
refinement requirement and then discharge them by automated reasoning via theorem proving.
Most of them support both algorithm refinement (refining a program fragment or an entire method into
an implementation, as we focus on in this paper),
and data refinement (refining abstract data in specification to concrete data in implementation).

Robin~\cite{Abrial:2010:ROT:1895404.1895406} is an open toolset which
integrates construction and  verification of Event-B Models.
The system behavior in Event-B is modeled by
action systems, i.e. a collection of variables and
guarded actions.
Abstract specification of a model is
constructed and then refined, following an incremental approach, however
object-oriented programs are not directly supported.

ProofPower Z~\cite{Imperial94ztools} is a tool interfaced with the YSE Zeta tool, which
supports refinement from Z specification to Ada programs. The tool
produces verification conditions for each refinement step for input into
a theorem prover, and produces Ada code by applying
the refinement steps. However, ProofPower and YSE are using different languages, making the communication
sometimes difficult. Also,
the un-readable output of proof in
ProofPower provides little guideline for locating failures in source programs.
Based on ProofPower Z,  the authors of~\cite{springerlink:10.1007/11901433_38,ZC11} develop
the automatic refinement of Circus, which is a refinement
language combining Z and CSP for describing state-rich reactive systems.

Perfect Developer~\cite{Crocker03perfectdeveloper:} is a software tool
for developing formal specifications and refining them into executable code.  Compared to
existing refinement tools before, it handles object-oriented features such as
inheritance, recursive call, polymorphism, dynamic binding, in conformance with behavioral sub-typing
principle. However, it does not support stepwise refinement, and requires a continuous strengthening of the code annotations, making the refinement less scalable.

Leino and Yessenov~\cite{Rustan10automatedstepwise} develop a refinement system
for object-oriented programs and
where the verification engine is based on the SMT solver Z3. It supports
automated stepwise refinement and all the intermediate steps are saved
using syntax of code skeletons during the whole refinement process.
This makes the location of failures in the source specification and code realizable.
Moreover, it handles aliasing between
data representations based on
the permission mechanism in Chalice~\cite{Leino09verificationof}.

An important strength of our work compared with these different approaches, in addition to the integration within the rCOS tool, a complete platform for software engineering, is the generation of a {\em proof witness}, {\em i.e.} we are not only answering if the refinement is correct or not, but also providing the reason why. Hence, we can easily reuse previously proved lemmas, making our approach more scalable.

%


\subsection{Proof Generation} 
\label{sub:proof_generation}

Our approach for proof generation was loosely inspired by the work realized with the automated demonstrator Zenon~\cite{DBLP:conf/lpar/BonichonDD07}. Indeed, Zenon can prove first order logical formulae using the tableau method, and generate the proof in Coq. It was originally developed for the Focalize~\cite{focalize} environment, which provides an expressive programming language where properties of a program can be proved in Coq, potentially using Zenon to generate some of the Coq proofs. An interesting feature is the definition of a simple and intuitive proof language following Lamport's guidelines~\cite{citeulike:791402}, which allows the user to break down a complex proof into small proofs, until a point where Zenon can prove automatically the statement. Such an integration is quite user-friendly, and therefore we aim at achieving a similar result.

In this context, the recent integration of Zenon with Isabelle and TLA+~\cite{DBLP:conf/cade/ChaudhuriDLM10} can probably be useful. We can also try to integrate our Maude module as an automatic theorem prover using the Sledgehammer in Isabelle~\cite{DBLP:journals/iandc/MengQP06}.



\section{Conclusion}
\label{sec:conclusion}

This paper presents, for any two programs defined within the rCOS tool, the generation of an Isabelle lemma stating that one program refines the other, as originally introduced in~\cite{LMW10}, and extends this approach by describing a Maude module that searches for a sequence of refinement steps, each step being implemented as a rewriting rule. If the Maude module can automatically find a sequence of rules, then it generates the Isabelle proof of the previous lemma, otherwise the lemma still needs to be manually proven.

The strengths of this approach are fourfold. Firstly, the generation process is integrated within the rCOS tool, and when the refinement can be automatically proven by Maude, then the process is transparent for the user. Secondly, the user has still the possibility to manually prove some lemmas, when Maude cannot automatically prove the refinement. Thirdly, new rules can be added to the process, simply by adding the new lemma in the Isabelle library and the new rewriting rule in the Maude file, without any need to modify existing code. Finally, it generates the witness of the proof, instead of only returning yes or no. It follows that any proven refinement can be stored, and re-used later, to prove more complex refinements. For instance, if we prove that the design of a method \rcos{foo} is refined by the design of a method \rcos{bar},
then we can re-use this proof of refinement in order to prove that a call to \rcos{foo} is refined by a call to \rcos{bar}.

This work mostly focuses on defining the framework where the different entities, {\em i.e.} the rCOS tool, Isabelle and Maude, can communicate together, in order to have a complete chain from the user of the rCOS tool, who is potentially an expert in software engineering rather than an expert in theorem-proving, to the proof of refinement in Isabelle. Hence, we have mostly considered simple examples, although rich enough to validate our approach, but clearly lacking the complexity of real-world programs.

As said in the Introduction, we build upon the previous encodings of the refinement calculus~\cite{Wright94,Laibinis00,Depasse01}, and as such, we do not present here complex programs, since the scalability problem is identical, and we prefer to focus here on the architecture of the framework. However, we believe that we have paved the way towards the certification of more complex programs, as we can leverage the incremental aspect of the refinement calculus. Indeed, a large, complex refinement chain, as it could be expected from a complex program, can always be decomposed as a sequence of simpler chains of refinement steps. Although some steps will probably always require a human interaction, such as the definition of a loop-invariant, some of tedious and repetitive steps can be automatically discharged.


A limitation of this approach is that, at this current stage, the Isabelle level depends on assumptions made at the rCOS level, in particular for the predicate
\isa{isGoodPath} and for the aliasing problem. Clearly, external tools to reason about the program structure to detect aliasing and cycle properties need
to be integrated in our approach.
In general, an interesting and challenging aspect of this work was to manage the fact that the programs we consider in Isabelle are generated from rCOS, and therefore comes with some implicit assumptions, such
as well-formedness and type safety. However, we have no way to {\em prove} these properties in Isabelle, and
we must limit ourselves to define and use well-formedness predicates. For instance, one of our previous
encodings was inconsistent due to the presence of axioms that were false for infinite graphs. Although
it was in practice impossible to generate an infinite graph from a correct rCOS program, we nevertheless had to
revise the encoding in order to remove these axioms.

As future work, following~\cite{Depasse01,Laibinis00}, more designs can be implemented in the translation process, such as recursive method calls.  Although rCOS supports dynamic binding, our encoding does not currently support it, but it could be done by adding a component to 
graph implementation to record actual types of objects, thus we can fix the method body actually called in the dynamic execution.

Finally, the Maude module can be optimized, in order to avoid generating proof-obligations that are clearly impossible, or in general to generate less proof-obligations. The use of Maude meta rules for rewrite strategies seems a suitable way to tackle this problem.






%



\end{document}